\documentclass[varenna]{cimento}

\usepackage{graphicx}  

\title{Symmetry and Supersymmetry in Nuclear Physics}

\author{A.~B. Balantekin}

\institute{Department of Physics, University of Wisconsin, Madison,
WI 53706, USA}

\shortauthor{A.~B. Balantekin}

\PACSes{\PACSit{03.65.Fd}{Algebraic methods}
\PACSit{21.60.Fw}{Models based on group theory}
\PACSit{25.70.Jj}{Fusion and fusion-fission reactions} 
\PACSit{26.30.-k}{Nucleosynthesis in novae, supernovae, and other 
explosive environments}}

\begin{document}

\maketitle

\begin{abstract}
A survey of algebraic approaches to various problems in nuclear physics is 
given. Examples are chosen from pairing of many-nucleon systems, nuclear 
structure, fusion reactions below the Coulomb barrier, and supernova 
neutrino physics to illustrate the utility of group-theoretical 
and related algebraic methods in 
nuclear physics. 
\end{abstract}

\section{Introduction}

Symmetry concepts play a very important role in all of physics. 
Originally symmetries utilized in physics did not change 
the particle statistics: such symmetries either transform bosons into
bosons or fermions into fermions. Natural mathematical tools to explore
symmetries are Lie algebras, i.e. the set of operators closing under 
commutation relations, schematically shown as  
\begin{equation}
\label{I1}
[ G_B, G_B ] = G_B .
\end{equation}
Lie algebras are associated with Lie groups. One could think of Lie groups 
as the exponentiation of Lie algebras. 

On the other hand, supersymmetries transform bosons into
bosons, fermions into fermions, {\it and} bosons into fermions and vice versa.
Natural mathematical tools to explore
them are superalgebras and supergroups:
Superalgebras, sets containing bosonic ($G_B$) as well as fermionic ($G_F$) 
operators, close under commutation {\it and} anticommutation relations as 
shown below:
\begin{eqnarray}
\label{I2}
                \left[ G_B, G_B \right] &=& G_B, \nonumber \\
                \left[ G_B, G_F \right] &=& G_F, \nonumber \\
                \{ G_F, G_F \} &=& G_B 
        \end{eqnarray}

The simplest superalgebra can be easily worked out. 
Consider three dimensional harmonic oscillator creation and annihilation
operators and define
\begin{equation}
\label{I3}
 K_0 = \frac{1}{2}\left( b_i^\dagger b_i+\frac{3}{2} \right)
\>\>\>\>\>
 K_+  = \frac{1}{2}b_i^\dagger b_i^\dagger = \left( K_-
                \right)^\dagger .
\end{equation}
It is easy to show that the operators defined in Eq. (\ref{I3}) satisfy  
\begin{equation}
\label{I4}
\left[ K_0,\ K_\pm \right] = \pm K_\pm, \>\>\>\>
 \left[ K_+,\ K_- \right] = - 2 K_0 .
\end{equation}
The algebra depicted in Eq. (\ref{I4}) is the SU(1,1) algebra 
\cite{Balantekin:1984hf}. Casimir operators are operators that 
satisfy the condition 
\begin{equation}
\label{I5}
    \left[ C(G),\ K_0 \right]=0=\left[ C(G),\ K_\pm \right]
\end{equation}
Casimir operators obtained by multiplying one, two, three
 elements of the algebra are called linear, quadratic,
cubic Casimir operators. For SU(1,1) the quadratic Casimir op. is
\begin{equation}
\label{I6}
 C_2=K_0^2-\frac{1}{2}\left( K_+K_-+K_-K_+ \right)
\end{equation}

The concept of dynamical symmetries found many applications in 
nuclear physics (see e.g. \cite{Iachello:2000ye}). 
Consider a chain of algebras (or associated groups):
\begin{equation}
\label{I7}
                G_1\supset G_2\supset\cdots\supset G_n
\end{equation}
        If a given Hamiltonian can be written in terms of the Casimir
        operators of the algebras in this chain, then such a Hamiltonian
        is said to possess a dynamical symmetry:
\begin{equation}
                H=\sum_{i=1}^{n}\left[ \alpha _iC_1(G_i)+\beta _iC_2(G_i)
        \right].
\end{equation}
Obviously all these Casimir
operators commute with each other.
Consequently, finding the energy eigenvalues of this Hamiltonian
reduces to the problem of reading the
eigenvalues of the Casimir operators off the existing tabulations.

Introducing spin (i.e., fermionic) degrees of freedom represented by the 
Pauli matrices as well as the bosonic
(harmonic oscillator) ones and defining 
\begin{equation}
 F_+ = \frac{1}{2}\sum_{i}^{}\sigma _ib_i^\dagger,\>\>\>\>\>
F_- = \frac{1}{2}\sum_{i}^{}\sigma _ib_i,
\end{equation}
one can write the following additional commutation and anticommutation 
relations:
\begin{eqnarray}
                \left[ K_0,\ F_\pm \right] &=& \pm\frac{1}{2}F_\pm,
\nonumber \\
                \left[ K_+,\ F_+ \right] &=& 0 = \left[ K_-,\ F_- \right],
\nonumber \\
                \left[ K_\pm,\ F_\mp \right] &=& \mp F_\pm, \nonumber \\
                \left\{ F_\pm,\ F_\pm \right\} &=& K_\pm, \nonumber \\
                \left\{ F_+,\ F_- \right\} &=& K_0 \nonumber
        \end{eqnarray}
The operators $K_+,\ K_-,\ K_0,\ F_+$, and $F_-$ generate
the Osp(1/2) superalgebra, which is non-compact 
\cite{Balantekin:1984hf,Balantekin:1987rp}. 
Since the operators $K_+,\ K_-,\ K_0$ alone generate the SP(2) 
$\sim$ SU(1,1) subalgebra we can write the group chain 
\begin{equation}
\label{Iz}
{\rm Osp}(1/2) \supset SU(1,1) \supset SO(2) .
\end{equation}
(Note that the operator $K_0$ can be viewed as the Casimir operator of 
the SO(2) subalgebra of SU(1,1)).  
The Casimir operators of Osp(1/2) and Sp(2) $\sim$ SU(1,1) are given by 
        \begin{eqnarray}
                C_2\left( {\rm Osp(1/2)} \right) &=& \frac{1}{4}\left( {\bf L}+
                \frac{\bf \sigma }{2} \right)^2 = \frac{1}{4}{\bf J}^2,
 \\
                C_2\left( {\rm Sp(2)} \right) &=& \frac{1}{2} {\bf L}^2 -
                \frac{3}{16},  
        \end{eqnarray} 
where ${\bf L}$ is the angular momentum carried by the oscillator, i.e. 
${\bf L}_i = \epsilon_{ijk}{\bf r}_j {\bf p}_k = i \epsilon_{ijk} 
b^{\dagger}_k b_j$. 
It can easily be shown that a harmonic oscillator Hamiltonian with a 
constant spin-orbit coupling 
\begin{equation}
\label{Ia}
 H = \frac{1}{2}\left( {\bf p}^2 + {\bf r}^2 \right)+\lambda
\left( {\bf \sigma} \cdot {\bf L} + \frac{3}{2} \right)
\end{equation}
can be written in terms of the Casimir operators of the group chain given in 
Eq. (\ref{Iz}): 
\begin{equation}
\label{Ib} 
H = 4\lambda C_2\left( {\rm Osp(1/2)} \right)-4\lambda C_2\left(
        {\rm Sp(2)} \right)+2K_0 .
\end{equation}
This provides perhaps the simplest example of a dynamical supersymmetry.

\section{Fermion pairing}

Pairing is a salient property of multi-fermion systems and as such it has 
a long history in nuclear physics (for a recent review see 
\cite{Dean:2002zx}).
Already in 1950 Meyer suggested that short-range attractive 
nucleon-nucleon interaction yields nuclear ground states with angular 
momentum zero \cite{Mayer}. 
Mean field calculations with effective interactions describe many
nuclear properties, however they cannot provide a complete solution 
of the underlying complex many-body problem. For example, 
after many years of investigations, we now that the structure of
low-lying collective states in medium-heavy to heavy nuclei are determined
by pairing correlations with L=0 and L=2. This was exploited by many
successful models of nuclear structure such as the Interacting Boson Model 
of Arima and Iachello \cite{Armia:1976ky,Arima:1978ha,Arima:1979zz}. 

Pairing plays a significant role not only in finite nuclei, but also in 
nuclear matter and can directly effect related observables. For example,  
we know that neutron superfluidity
is present in the crust and the inner part of a neutron star. Pairing
could significantly effect the thermal evolution of the neutron star by
suppressing neutrino (and possibly exotics such as axions) emission 
\cite{Page:2000wt}. 

Charge symmetry implies that interactions between
two protons and two neutrons are very similar; hence proton-proton and 
neutron-neutron pairs play a similar role in nuclei. 
In addition isospin symmetry implies that proton-neutron interaction
is also very similar to the proton-proton and neutron-neutron 
interactions. Currently 
there is very little experimental information about 
neutron-proton pairing in heavier nuclei as one needs to study proton 
rich nuclei to achieve this goal. But one expects that data from 
the current and future radiative beam facilities will change this picture.

First microscopic theory of pairing was the Bardeen, Cooper, Schriffer (BCS)
theory \cite{Bardeen:1957mv}.
Soon after its introduction, the BCS theory was applied to nuclear structure
\cite{bohr,belyaev,migdal}. 
Application of the BCS theory to nuclear structure has a main drawback:
BCS wave function is {\it not} an eigenstate of the number operator. Several
solutions were offered to remedy this shortcoming such as 
adding Random-Phase Approximation (RPA) to the BCS theory
\cite{unna}, projection of the particle 
number after variation \cite{kerman1}, or 
projection of the particle number before the
variation. The last technique was recently much utilized 
in nuclei (see e.g. \cite{Hagino:2000ba}). 
Pairing correlations may also play an interesting role in halo nuclei 
\cite{Hagino:2006ib}. 
It should also be noted that the theory of pairing in nuclear physics has many 
parallels with the theory of ultrasmall metallic grains in condensed 
matter physics (see e.g. \cite{vonDelft:1999si}). 

\subsection{Quasi-Spin Algebra}

The concept of seniority was introduced by Racah to aid the classification 
of atomic spectra \cite{racah}. 
Seniority quantum number is basically the number of unpaired particles 
in the $j^n$ configuration. Seniority-conserving pairing interactions are 
a very limited class, however such interactions make an interesting case 
study. Kerman introduced the quasi-spin scheme to treat such cases 
\cite{kerman2}.  In this scheme nucleons are placed at time-reversed 
states $|j \>
m \rangle$ and $(-1)^{(j-m)}|j \> -m\rangle$. Introducing the creation and
annihilation operators for nucleons at level $j$,  $a^\dagger_{j\>m}$ 
and $a_{j\>m}$, the quasi-spin operators are written as 
\begin{equation}
\hat{S}^+_j = \sum_{m>0} (-1)^{(j-m)} a^\dagger_{j\>m}a^\dagger_{j\>-m},
\end{equation}
\begin{equation}
\hat{S}^-_j = \sum_{m>0} (-1)^{(j-m)} a_{j\>-m}a_{j\>m}, 
\end{equation}
and
\begin{equation}
\hat{S}^0_j=\frac{1}{2}\sum_{m>0}
\left(a^\dagger_{j\>m}a_{j\>m}+a^\dagger_{j\>-m}a_{j\>-m}-1,
\right)
\end{equation}
These operators form a set of mutually commuting SU(2) algebras:
\begin{equation}
[\hat{S}^+_i, \hat{S}^-_j ] = 2 \delta_{ij} \hat{S}^0_j, \>\>\>\>\>\>\>
[\hat{S}^0_i, \hat{S}^{\pm}_j] = \pm \delta_{ij} \hat{S}^{\pm}_j. 
\end{equation} 
The operator $\hat{S}^0_j$ can be related to the number operator  
\begin{equation}
\hat{S}^0_j=\hat{N}_{j}-\frac{1}{2}\Omega_j, 
\end{equation} 
where $\Omega_j=j+\frac{1}{2}$ is the maximum number of pairs that
can occupy the level $j$ and the number operator is 
\begin{equation}
\hat{N}_j=\frac{1}{2}\sum_{m>0}
\left(a^\dagger_{j\>m}a_{j\>m}+a^\dagger_{j\>-m}a_{j\>-m}\right).
\end{equation}
Since $0 < \hat{N}_j <\Omega_j$ these SU(2) algebras are realized in the 
representation with the total angular momentum quantum number  
$\frac{1}{2}\Omega_j$. 

The most general Hamiltonian for nucleons interacting with a pairing force 
can be written as 
\begin{equation}
\label{20}
\hat{H}=\sum_{jm} \epsilon_j a^\dagger_{j\>m} a_{j\>m} -
|G|\sum_{jj'}c_{jj'} \hat{S}^+_j \hat{S}^-_{j'}, 
\end{equation}
where  $\epsilon_j$ are the single
particle energy levels, $|G|$ is the pairing interaction strength with 
dimensions of energy, and $c_{jj'}$ are dimensionless parameters describing 
the distribution of this strength between different orbitals.  
When the latter are separable ($c_{jj'}=c^*_jc_{j'}$) we get 
\begin{equation}
\label{21}
\hat{H}= \sum_{jm} \epsilon_j a^\dagger_{j\>m} a_{j\>m} -
|G|\sum_{jj'}c^*_jc_{j'} \hat{S}^+_j \hat{S}^-_{j'}.
\end{equation}
There are a number of approximations one can make to simplify the 
Hamiltonians in Eqs. (\ref{20}) or (\ref{21}). 
If we assume that the NN interaction is determined by a single
parameter (usually chosen to be the scattering length), 
all $c_j$'s are the same and we get
\begin{equation}
\hat{H}=\sum_{jm} \epsilon_j a^\dagger_{j\>m} a_{j\>m} -
|G|\sum_{jj'} \hat{S}^+_j \hat{S}^-_{j'}.
\end{equation}
This case was solved by Richardson \cite{rich}. 

If we assume that the energy levels are degenerate then  
the first term is a constant for a given fixed number of pairs.
This case can be solved by using the quasispin algebra 
since $H \propto S^+ S^-$. A list of the exactly solvable cases 
can then be given as follows: 
\begin{itemize}
\item Quasi-spin limit \cite{kerman2}: 
\begin{equation}
\hat{H}=- |G|\sum_{jj'}  \hat{S}^+_j \hat{S}^-_{j'}.
\end{equation}
\item Richardson's limit \cite{rich}:
\begin{equation}
\label{25} 
\hat{H}=\sum_{jm} \epsilon_j a^\dagger_{j\>m} a_{j\>m} -
|G|\sum_{jj'} \hat{S}^+_j \hat{S}^-_{j'}.
\end{equation}
\item An algebraic solution developed by Gaudin \cite{gaudin} 
which is sketched out in the next section (section \ref{gaudinsection}). 
Gaudin's model is closely related to Richardson's limit. 
\item The limit in which the energy levels are degenerate
(when the first term becomes  a constant for a given number of pairs):
\begin{equation}
\hat{H}=- |G|\sum_{jj'}c^*_jc_{j'} \hat{S}^+_j \hat{S}^-_{j'}.
\end{equation} 
Different aspects of this solution was worked out in Refs. 
\cite{Pan:1997rw,Balantekin:2007vs,Balantekin:2007qr}.
\item 
Most general separable case with two shells \cite{Balantekin:2007ip}.  
\end{itemize}
Clearly we can solve the pairing problem numerically in the
quasispin basis \cite{Volya:2000ne}. Note that it is also possible to utilize 
the quasi-spin concept for mixed systems of bosons and fermions in a  
supersymmetric framework \cite{Schmitt:1998qk}.

\subsection{Gaudin Algebra}
\label{gaudinsection}

To study the problem of interacting spins on a lattice 
Gaudin introduced a method \cite{gaudin}  
closely related to the Richardson's solution. 
Our presentation of Gaudin's method here follows that given in Ref. 
\cite{Balantekin:2005sj}. This method starts with three 
operators $J^{\pm}(\lambda), J^0(\lambda)$, parametrized by one 
parameter $\lambda$, satisfying the commutation relations 
\begin{equation}
\label{27}
[J^+(\lambda),J^-(\mu)]=2\frac{J^0(\lambda)-J^0(\mu)}{\lambda-\mu},
\end{equation}
\begin{equation}
\label{28}
[J^0(\lambda),J^{\pm}(\mu)]=\pm\frac{J^{\pm}(\lambda)-J^{\pm}(\mu)}
                                                  {\lambda-\mu},
\end{equation}
and 
\begin{equation}
\label{29}
[J^0(\lambda),J^0(\mu)]=[J^{\pm}(\lambda),J^{\pm}(\mu)]=0. 
\end{equation}
The Lie algebra depicted above is referred to as the rational Gaudin algebra. 
A possible realization of this algebra is given by 
\begin{equation}
\label{30}
J^{0}(\lambda)=\sum_{i=1}^N\frac{\hat{S}^0_i}{\epsilon_i-\lambda}
\quad\mbox{and}\quad
J^{\pm}(\lambda)=\sum_{i=1}^N \frac{\hat{S}^{\pm}_i}{\epsilon_i-\lambda}, 
\end{equation}
where $\epsilon_i$ are, in general, arbitrary parameters. In Eq. (\ref{30}), 
instead of the quasi-spin algebra one can 
obviously use any mutually-commuting $N$ SU(2) 
algebras (in fact, copies of any other algebra). The operator  
\begin{equation}
\label{31}
H(\lambda)=J^0(\lambda)J^0(\lambda)+\frac{1}{2}J^+(\lambda)J^-(\lambda)+
           \frac{1}{2}J^-(\lambda)J^+(\lambda)
\end{equation} 
is not the Casimir operator of the Gaudin algebra, but such operators 
commute for different values of the parameter:  
\begin{equation}
\label{32}
[H(\lambda),H(\mu)]=0 . 
\end{equation}                                         
Lowest weight vector, $|0 \rangle$, is chosen to satisfy the conditions 
\begin{equation}
J^-(\lambda)|0 \rangle =0,\quad\mbox{and}\quad
J^0(\lambda)|0 \rangle =W(\lambda)|0 \rangle.
\end{equation}
Hence it is an eigenstate of the Hamiltonian given in Eq. (\ref{31}): 
\begin{equation}
H(\lambda)|0 \rangle = \left[ W(\lambda)^2-W'(\lambda) \right] |0 \rangle .
\end{equation} 
To find other eigenstates consider the state $|\xi \rangle
\equiv J^+(\xi)|0 \rangle$ for an arbitrary complex number
$\xi$. Since
\begin{equation}
[H(\lambda),J^+(\xi)]=\frac{2}{\lambda-\xi}\left(J^+(\lambda)J^0(\xi)
-J^+(\xi)J^0(\lambda)\right), 
\end{equation}
we conclude that if $W(\xi)=0$, then $J^+(\xi)|0 \rangle$ is an eigenstate of
$H(\lambda)$ with the eigenvalue
\begin{equation}
E_1(\lambda)= \left[ W(\lambda)^2-W'(\lambda) \right]
-2\frac{W(\lambda)}{\lambda-\xi}.
\end{equation}
Gaudin showed that this approach can be generalized and a state of the form
\begin{equation}
|\xi_1,\xi_2,\dots,\xi_n>\equiv J^+(\xi_1) J^+(\xi_2)\dots
J^+(\xi_n)|0>
\end{equation}
is an eigenvector of $H(\lambda)$ if the numbers
$\xi_1,\xi_2,\dots, \xi_n\in {\bf{C}}$ satisfy the so-called Bethe
Ansatz equations:
\begin{equation}
W(\xi_\alpha)=\sum_{ {\beta=1}\atop{(\beta\neq\alpha)} }^n
\frac{1}{\xi_\alpha-\xi_\beta} \quad \mbox{for} \quad
\alpha=1,2,\dots,n.
\end{equation}
Corresponding eigenvalue is
\begin{equation}
E_n(\lambda) = \left[ W(\lambda)^2-W'(\lambda) \right]
-2\sum_{\alpha=1}^n
\frac{W(\lambda)-W(\xi_\alpha)}{\lambda-\xi_\alpha}.
\end{equation}
                                                               
To make a connection to Richardson's solution we define so-called 
${\cal R}$-operators as 
\begin{equation}
\label{40}
\lim_{\lambda \rightarrow \epsilon_k} (\lambda - \epsilon_k) H(\lambda) =
{\cal R}_k.
\end{equation}
In the realization of Eq. (\ref{30}), these operators take the form 
\begin{equation}
{\cal R}_k = -2 \sum_{j\neq k} \frac{{\bf S}_k \cdot {\bf S}_j}{\epsilon_k -
\epsilon_j}.
\end{equation}
Taking the limits $\mu \rightarrow \epsilon_k$ first and 
$\lambda \rightarrow \epsilon_j$ second in Eq. (\ref{32}), one 
easily obtains 
\begin{equation}
\label{42}
[H(\lambda), {\cal R}_k] = 0, \>\>\>
[{\cal R}_j, {\cal R}_k] = 0 .
\end{equation}
One can also prove the equalities 
\begin{equation}
\sum_i {\cal R}_i =0 ,
\end{equation}
and
\begin{equation}
\sum_i \epsilon_i {\cal R}_i = - 2 \sum_{i \neq j} {\bf S}_i \cdot {\bf S}_j .
\end{equation}

A careful examination of the Gaudin algebra in Eqs. (\ref{27}), (\ref{28}), 
and (\ref{29}) indicates that 
not only the operators ${\bf J} (\lambda)$, but also
the operators ${\bf J} (\lambda) + {\bf c}$ satisfy this algebra for
a constant ${\bf c}$. In this case the conserved quantity is replaced by 
\begin{equation}
\label{45}
H(\lambda) = {\bf J}(\lambda) \cdot {\bf J}(\lambda) \Rightarrow
H(\lambda) + 2 {\bf c} \cdot {\bf J}(\lambda) + {\bf c}^2
\end{equation}
which has the same eigenstates. One can define Richardson operators, $R_k$, 
in an analogous way to the ${\cal R}_k$ defined in Eq. (\ref{40}): 
\begin{equation}
\label{46}
\lim_{\lambda \rightarrow \epsilon_k} (\lambda - \epsilon_k)
\left( H(\lambda)
+ 2 {\bf c} \cdot {\bf S} \right) =
R_k, 
\end{equation}
which implies 
\begin{equation}
R_k = - 2 {\bf c} \cdot {\bf S}_k -
2 \sum_{j\neq k} \frac{{\bf S}_k \cdot {\bf S}_j}{\epsilon_k -
\epsilon_j}. 
\end{equation}
Eq. (\ref{42}) is then replaced by 
\begin{equation}
\label{48}
[H(\lambda) + 2 {\bf c} \cdot {\bf S}, R_k]=0 \>\>\>\>\>
[R_j,  R_k] =0, 
\end{equation}
with the conditions 
\begin{equation}
\sum_i R_i = - 2  {\bf c} \cdot \sum_k {\bf S}_k, 
\end{equation}
and 
\begin{equation}
\sum_i \epsilon_i R_i = -2 \sum_i\epsilon_i  {\bf c} \cdot {\bf S}_i - 2
\sum_{i \neq j} {\bf S}_i \cdot {\bf S}_j .
\end{equation}

Rewriting the Hamiltonian of Eq. (\ref{25}) in the form 
\begin{eqnarray}
\Rightarrow H &=&\sum_j \epsilon_j S^0_j - |G| \left( \left( \sum_i {\bf S}_i 
\right) \cdot \left( \sum_i {\bf S}_i \right) - \left( \sum_i S^0_i \right)^2 
+ \left( \sum_i S^0_i \right)
\right)  \\ &+& {\rm constant \> terms},  \nonumber 
\end{eqnarray} 
and choosing the constant vector of Eq. (\ref{45}) to be 
\begin{equation}
{\bf c} = (0, 0, -1/2|G|)
\end{equation}
one immediately obtains 
\begin{equation}
\label{53}
\frac{H}{|G|} = \sum_i \epsilon_i R_i + |G|^2 \left( \sum_i R_i \right)^2 
- |G| \sum_i R_i
+ \cdots
\end{equation}
Since all $R_k$ and $H(\lambda) + 2 {\bf c} \cdot {\bf S}$ mutually 
commute (cf. Eq. (\ref{48})), they 
have the same eigenvalues. Eq. (\ref{53}) then tells us that they are also 
the eigenvalues of the Richardson Hamiltonian, Eq. (\ref{25}). 
 
\subsection{Exact Solution for Degenerate Spectra} 

If the single particle spectrum is degenerate, then it is possible to find 
the eigenvalues and eigenstates of the Hamiltonian in Eq. (\ref{21}). 
If all the single particle energies are the same, the first term in Eq. 
(\ref{21}) is a constant and can be ignored.  
Defining the operators 
\begin{equation}
\label{54} 
\hat{S}^+(0)=\sum_j c^*_j\hat{S}^+_j \ \ \ \ \mbox{and} \ \ \ \ \
\hat{S}^-(0)=\sum_j c_j\hat{S}^-_j,
\end{equation}
the Hamiltonian of Eq. (\ref{21}) can be rewritten as 
\begin{equation}
\label{55}
\hat{H}=-|G|\hat{S}^+(0)\hat{S}^-(0) + {\rm constant}. 
\end{equation}
(The operators in Eq. (\ref{54}) are defined with argument $0$ for reasons 
explained in the following discussion).  
In the 1970's Talmi showed that,  
under certain assumptions, a state of the form
\begin{equation}
\hat{S}^+(0)|0\rangle=\sum_j c^*_j\hat{S}^+_j |0\rangle , 
\end{equation}
with $|0\rangle$ being the particle vacuum, 
is an eigenstate of a class of Hamiltonians including the one above 
\cite{talmi}. A direct calculation yields the eigenvalue equation 
\begin{equation}
\label{57} 
\hat{H}\hat{S}^+(0)|0\rangle= \left( -|G|\sum_j \Omega_j |c_j|^2 \right)
\hat{S}^+(0)|0\rangle .
\end{equation}
However, there are other one-pair states besides 
the one in Eq. (\ref{57}). For example for two levels
$j_1$ and $j_2$, the orthogonal state
\begin{equation}
\left( \frac{c_{j_2}}{\Omega_{j_1}} \hat{S}^+_{j_1} -
\frac{c_{j_1}}{\Omega_{j_2}} \hat{S}^+_{j_2} \right) |0 \rangle,
\end{equation}
is also an eigenstate with E=0. In Ref. \cite{Pan:1997rw}
it was shown that there is a systematic way to derive these states. 
To see their solution we define the operators 
\begin{equation}
\label{59}
\hat{S}^+(x)=\sum_j\frac{c^*_j}{1-|c_j|^2x}\hat{S}^+_j \ \ \ \ \
\mbox{and} \ \ \ \ \
\hat{S}^-(x)=\sum_j\frac{c_j}{1-|c_j|^2x}\hat{S}^-_j.
\end{equation}
Note that if one substitutes $x=0$ in the operators of Eq. (\ref{59}), 
one obtains the operators in Eq. (\ref{54}). Further defining the operator 
\begin{equation}
\hat{K}^0(x)=\sum_j\frac{1}{1/|c_j|^2-x}\hat{S}_j^0 ,
\end{equation}
one can prove the following commutation relations: 
\begin{equation}
\label{61}
[\hat{S}^+(x),\hat{S}^-(0)] = [\hat{S}^+(0),\hat{S}^-(x)] =
2K^0(x)
\end{equation}
\begin{equation}
\label{62}
[\hat{K}^0(x),\hat{S}^\pm(y)] = \pm
\frac{\hat{S}^\pm(x)-\hat{S}^\pm(y)}{x-y}
\end{equation}
These commutators are very similar to, but not the same as, those of the  
Gaudin algebra described in Eqs. (\ref{27}), (\ref{28}), and (\ref{29}). 
Using the commutators in Eqs. (\ref{61}) and (\ref{62})
one can easily show that 
\begin{equation}
\label{63}
\hat{S}^+(0)\hat{S}^+(z^{(N)}_1) \dots
\hat{S}^+(z^{(N)}_{N-1})|0\rangle
\end{equation}
is an eigenstate of the Hamiltonian in Eq. (\ref{55}) 
if the following Bethe ansatz equations
are satisfied:
\begin{equation}
\label{64}
\sum_j \frac{-\Omega_j/2}{1/|c_j|^2-z^{(N)}_m}
=\frac{1}{z^{(N)}_m}+\sum_{k=1(k\neq m)}^{N-1}
\frac{1}{z^{(N)}_m-z^{(N)}_k} \ \ \ \ \ \ \
m=1,2,\dots N-1.
\end{equation}
The energy of the state in Eq. (\ref{63}) is 
\begin{equation}
\label{65}
E_N =-|G|\left(\sum_j \Omega_j |c_j|^2-\sum_{k=1}^{N-1}
\frac{2}{z^{(N)}_k}\right) .
\end{equation} 
The authors of Ref. \cite{Pan:1997rw} used a Laurent expansion of the 
operators given in Eq. (\ref{59})
around $x=0$ followed by an analytic continuation argument to
show the validity of their results in the entire complex
plane except some singular points. The derivation sketched above 
instead utilizes the algebra depicted in Eqs. (\ref{61}) and (\ref{62}); 
it is significantly simpler.  
The state in Eq. (\ref{63}) is an eigenstate if the shell is at most half
full. Similarly the state 
\begin{equation}
\hat{S}^+(x^{(N)}_1)\hat{S}^+(x^{(N)}_2) \dots
\hat{S}^+(x^{(N)}_N)|0\rangle
\end{equation}
is an eigenstate with zero energy if the following Bethe ansatz equations
are satisfied:
\begin{equation}
\label{67} 
\sum_j \frac{-\Omega_j/2}{1/|c_j|^2-x^{(N)}_m}=\sum_{k=1(k\neq
m)}^N \frac{1}{x^{(N)}_m-x^{(N)}_k} \ \ \ \ \ \ \mbox{for every} \
\ \ m=1,2,\dots,N
\end{equation}
Again this is an eigenstate if the shell is at most half
full.

To figure out what happens if the available states are more than half full 
we note that there are degeneracies in the spectra. Let us denote the 
state where all levels are completely filled by $|\bar{0}\rangle$. 
It is easy to show that the state $|\bar{0}\rangle$ and the particle vacuum, 
$|0\rangle$ are both eigenstates of the Hamiltonian of Eq. (\ref{55}) with 
the same energy, $E = - |G| \sum_j \Omega_{j} |c_j|^2$. 
This suggests that if the shells are more than half full then  
one should start with a state of the form 
\begin{equation}
\label{68}
\hat{S}^-(z_1^{(N)})\hat{S}^-(z_2^{(N)})\dots\hat{S}^-(z_{N-1}^{(N)})
|\bar{0}\rangle .
\end{equation}
Indeed the state in Eq. (\ref{68}) 
is an eigenstate of the Hamiltonian in Eq. (\ref{55}) 
with the energy 
\begin{equation}
\label{69}
E = -G \left( \sum_j \Omega_j |c_j|^2-\sum_{k=1}^{N-1}
\frac{2}{z^{(N)}_k} \right)
\end{equation}
if the following Bethe ansatz equations are satisfied 
\cite{Balantekin:2007vs,Balantekin:2007qr}:  
\begin{equation}
\label{70}
\sum_j
\frac{-\Omega_j/2}{1/|c_j|^2-z^{(N)}_m}
=\frac{1}{z^{(N)}_m}+\sum_{k=1(k\neq m)}^{N-1}
\frac{1}{z^{(N)}_m-z^{(N)}_k} .
\end{equation}
In Eq. (\ref{70}) $N_{max}+1-N$ is the number of particle pairs. 
If the available 
particle states are more than half full, then there are no zero energy 
states. 
Note that the Bethe ansatz conditions in Eqs. (\ref{70}) and 
(\ref{64}) as well the energies in Eqs. (\ref{69}) and (\ref{65}) 
are identical. This particle-hole degeneracy is due to a hidden supersymmetry 
\cite{Balantekin:2007qr} and is illustrated in Table \ref{Table1}. 
This supersymmetry is described in Section \ref{susyofpair}.

\begin{table}
\caption{Particle-hole degeneracy - two different states with the same energy} 
\label{Table1}
\begin{tabular}{rl}
\hline  
 {\bf No. of Pairs} & {\bf State} \\
\hline 
$N$   &
$ \hat{S}^+(0)\hat{S}^+(z^{(N)}_1) \dots
\hat{S}^+(z^{(N)}_{N-1})|0\rangle $
 \\
\hline 
$N_{max} + 1 -N$ &
$\hat{S}^-(z_1^{(N)})\hat{S}^-(z_2^{(N)})\dots\hat{S}^-(z_{N-1}^{(N)})
|\bar{0}\rangle $
\\
\hline
\end{tabular}
\end{table}

\subsection{Exact Solutions with two shells}

Consider the most general pairing Hamiltonian with only two shells:
\begin{equation}
\label{71}
\frac{\hat{H}}{|G|}= \sum_{j} 2\varepsilon_j
\hat{S}_j^0 - \sum_{jj'}c^*_jc_{j'} \hat{S}^+_j
\hat{S}^-_{j'}+\sum_j\varepsilon_j\Omega_j,
\end{equation}
with $\varepsilon_j=\epsilon_j/|G|$. It turns out that this problem can be 
solved using techniques illustrated in the previous sections. 
Eigenstates of the Hamiltonian in Eq. (\ref{71}) can be written using the 
new step operators \cite{Balantekin:2007ip}
\begin{equation}
{\cal J}^+(x)=\sum_j\frac{c_j^*}{2\varepsilon_j-|c_j|^2x}S_j^+
\end{equation}
as
\begin{equation}
{\cal J}^+(x_1){\cal J}^+(x_2)\cdots {\cal J}^+(x_N)|0\rangle .
\end{equation}
Introducing the definitions 
\begin{equation}
\beta=2\frac{\varepsilon_{j_1}-\varepsilon_{j_2}}{|c_{j_1}|^2-|c_{j_2}|^2}
\quad \quad \quad  \delta
=2\frac{\varepsilon_{j_2}|c_{j_1}|^2-\varepsilon_{j_1}|c_{j_2}|^2}
{|c_{j_1}|^2-|c_{j_2}|^2}.
\end{equation}
we obtain the energy eigenvalue to be 
\begin{equation}
E_N=-\sum_{n=1}^N\frac{\delta x_n}{\beta-x_n} 
\end{equation}
if the parameters $x_k$ satisfy
the Bethe ansatz equations
\begin{equation}
\sum_{j}\frac{\Omega_j|c_j|^2}{2\varepsilon_j-|c_{j}|^2x_k}
=\frac{\beta}{\beta-x_k} +\sum_{n=1(\neq k)}^N\frac{2}{x_n-x_k}.
\end{equation}

\subsection{Solutions of Bethe Ansatz equations}

Various solutions of the pairing problem discussed in the previous sections 
are typically considered as semi-analytical solutions since one still needs 
to find the solutions of the Bethe ansatz equations. This a task which, 
quite often, needs to be tackled numerically. However, in certain limits it 
is possible to find solutions of the Bethe ansatz equations analytically.
The method outlined here was first presented in Ref. \cite{Balantekin:2004yf}. 

Consider the Bethe ansatz equations for the degenerate single particle 
levels and zero energy eigenstate given in Eq. (\ref{67}). Introducing 
new variables, $\eta_i$, 
\begin{equation}
x_{i}^{(N)}=\frac{1}{|c_{j_2}|^2} +
\eta_i^{(N)}\left(\frac{1}{|c_{j_1}|^2}-\frac{
1}{|c_{j_2}|^2}\right)
\end{equation}
Eq. (\ref{67}) can be rewritten as 
\begin{equation}
\label{78}
\sum_{k=1(k\neq i)}^{N}\frac{1}{\eta_{i}^{(N)}-\eta_{k}^{(N)}}
-\frac{\Omega_{j_2}/2}{\eta_i^{(N)}}+\frac{\Omega_{j_1}/2}{1-\eta_i^{(N)}} 
= 0 .
\end{equation}
It can be easily shown that the polynomial admitting the solutions of 
Eq. (\ref{78}) as zeros 
\begin{equation}
p_N(z) = \prod_{i =1}^{N} (z - \eta_i^{(N)})
\end{equation}
satisfies the hypergeometric equation \cite{stiel}
\begin{equation}
z(1-z) p_{N}^{\prime \prime }
+ \left[-\Omega_{j_2}+\left(\Omega_{j_1}  \Omega
_{j_2}\right)z\right] p_{N}^{\prime } +
N\left(N-\Omega_{j_1}-\Omega_{j_2}-1\right) p_N=0 . 
\end{equation}
Consequently the problem of finding the solutions of the Bethe ansatz 
equation (\ref{67}) reduces to calculating the roots of hypergeometric 
functions. For analytical expressions of the energy eigenvalues 
obtained in this manner the reader is referred to 
Ref \cite{Balantekin:2007vs}. 

\section{Supersymmetric Quantum Mechanics in Nuclear Physics}

Consider two Hamiltonians
\begin{equation}
\label{81}
        H_1=G^\dagger G,\ H_2=GG^\dagger,
\end{equation}
where $G$ is an arbitrary operator. The
eigenvalues of these two Hamiltonians
        \begin{eqnarray}
\label{susyqm}
                G^\dagger G|1,n\rangle &=& E_n^{(1)}|1,n\rangle ,
 \\
                \label{eq16b}GG^\dagger|2,n\rangle &=& E_n^{(2)}|2,n\rangle
\nonumber
\end{eqnarray}
are the same:
\begin{equation}E_n^{(1)}=E_n^{(2)}=E_n
\end{equation}
and the eigenvectors are related:
\begin{equation}
        |2,n\rangle=G\left[ G^\dagger G \right]^{-1/2}|1,n\rangle.
\end{equation}
This works for all cases except when
$G|1,n\rangle=0$, which should be the ground state energy of the
positive-definite Hamiltonian $H_1$.

To see why this is called supersymmetry we define
\begin{equation}
        Q^\dagger=\left(
        \begin{array}{cc}
                0 & 0\\
                G^\dagger & 0
        \end{array}
         \right),\quad Q=\left(
         \begin{array}{cc}
                0 & G\\
                0 & 0
         \end{array}
          \right),
\end{equation}
Then
\begin{equation}
\label{85}
        H=\left\{ Q,Q^\dagger \right\}=\left(
        \begin{array}{cc}
                H_2 &0\\
                0 & H_1
        \end{array}
         \right).
\end{equation}
with
\begin{equation}
\label{86}
[H,Q]=0=[H,Q^\dagger].
\end{equation}
Clearly the operators $H$, $Q$, and $Q^{\dagger}$ close under the commutation 
and anticommutation relations of Eq. (\ref{85}) and (\ref{86}), forming a 
very simple superalgebra. The two Hamiltonians depicted in Eq. (\ref{81}) 
are said to form a system of supersymmetric quantum mechanics 
\cite{Witten:1981nf}. Many aspects of the supersymmetric 
quantum mechanics has been investigated 
in detail \cite{Cooper:1994eh,Cooper:1986tz,Fricke:1987ft}. 
In the following sections two applications of supersymmetric quantum mechanics 
to nuclear physics are summarized. 

\subsection{Application of Supersymmetric Quantum Mechanics to 
Pseudo-Orbital Angular Momentum and Pseudospin}

The nuclear shell model is a mean-field theory where the single
particle levels can be taken as those of a three-dimensional 
harmonic oscillator (labeled with SU(3) quantum numbers) for the lowest 
($A \le 20$) levels. For heavier nuclei with more
than 20 protons or neutrons, different parity orbitals mix.
The Nilsson Hamiltonian of the spherical shell model is 
\cite{nilsson} 
\begin{equation}
\label{92}
        H=\omega b_i^\dagger b_i - 2k {\bf L.S} - k\mu {\bf L}^2,
\end{equation}
where the second term mixes opposite parity orbitals
and the last term mocks up the deeper potential felt by the
nucleons as $L$ increases.

Fits to data indicate that $\mu \approx 0.5$, hence there 
exists degeneracies in the single
particle spectra. In the 50--82 shell (the SU(3)
label or the principal harmonic oscillator quantum number of which is
$N=4$) the s$_{1/2}$ and d$_{3/2}$ orbitals and further d$_{5/2}$
and g$_{7/2}$ orbitals are almost degenerate.
It is possible to
give a phenomenological account of this degeneracy by introducing
a second SU(3) algebra called the pseudo-SU(3) \cite{pse1,pse2}. 
Assuming that 
those orbitals belong to the $N=3$ (with $\ell=1,3$)
representation of the latter $SU(3)$ algebra, 
the quantum numbers of the SO(3) algebra included in this new
SU(3) are called pseudo-orbital-angular momentum ($\ell=1,3$ in this
case). We can also introduce a ``pseudo-spin'' ($s=\frac{1}{2}$).
One can easily show that $j=1/2$ and 3/2 orbitals (and also $j=5/2$
and 7/2 orbitals) are degenerate if pseudo-orbital angular
momentum and pseudo-spin coupling vanishes. This degeneracy follows 
form the supersymmetric quantum mechanical nature of the problem. 

It can be shown that two Hamiltonians written in the 
SU(3) and the pseudo-SU(3) bases are supersymmetric partners of each other 
\cite{Balantekin:1992qp}. 
The operator that transforms these two bases into one another is 
\cite{quesne}
\begin{eqnarray}
        U &=& G\left[ G^\dagger G \right]^{-1/2}=\sqrt{2}F_-\left(
        K_0+\left[ F_+,F_- \right] \right)^{-1/2} \\
         &=& \left( \sigma _ib_i^\dagger \right)\left( b_i^\dagger b_i -
         \sigma _iL_i \right)^{-1/2} \nonumber
\end{eqnarray}
yielding the supersymmetry transformation between the pseudo-SU(3) 
Hamiltonian $H'$ and the SU(3) Hamiltonian $H$:  
\begin{equation}   
\label{89}
        H' = U\ H U^\dagger  
         = b_i^\dagger b_i - 2k\left( 2\mu -1 \right) {\bf L \cdot  S} -
         k\mu {\bf L}^2+\left[ 1-2k(\mu -1) \right] .
\end{equation}
The transformation depicted in Eq. (\ref{89}) can also be expressed in 
terms of the generators of the  orthosymplectic superalgebra Osp(1/2)
\cite{Balantekin:1992qp}. 

\subsection{Supersymmetric Quantum Mechanics and Pairing in Nuclei}
\label{susyofpair}

Let us consider the separable pairing Hamiltonian with degenerate 
single-particle spectra given in Eq. (\ref{55}): 
\begin{equation} 
\label{91}
\hat{H}_{SC} \sim -|G|\hat{S}^+(0)\hat{S}^-(0),
\end{equation}
and introduce the operator
\begin{equation} 
\hat{T} = \exp \left( - i \frac{\pi}{2} \sum_i (\hat{S}_i^+ + \hat{S}_i^-)
\right)
\end{equation}
This operator transforms the empty shell, $|0\rangle$, to the fully
occupied shell,  
$|\bar{0}\rangle$:
\begin{equation} 
\hat{T} |0\rangle=|\bar{0}\rangle .
\end{equation}
To establish the connection to the supersymmetric quantum mechanics we 
define the operators 
\begin{equation}   
\hat{B}^- = \hat{T}^{\dagger} \hat{S}^-(0),\ \ \ \ \ \ \ \
\hat{B}^+ = \hat{S}^+(0) \hat{T}.
\end{equation}  
Supersymmetric quantum mechanics tells us that the partner Hamiltonians
$\hat{H}_1 = \hat{B}^+ \hat{B}^-$ and $\hat{H}_2 = \hat{B}^- \hat{B}^+$
have identical spectra except
for the ground state of $\hat{H}_1$. 
It can easily be shown that in this case two Hamiltonians
$\hat{H}_1$ and $\hat{H}_2$ are
actually identical and equal to the pairing Hamiltonian 
(Eq. (\ref{91})). Hence the role of
the supersymmetry is to connect the states
$| 1, n \rangle$ and $| 2, n \rangle$. 
These are the ``particle'' and ``hole'' states. 
Hence if the shell is less than 
half-full (``particles'') there is a zero-energy state (note that 
the Hamiltonian in Eq. (\ref{55}) negative definite, hence this is the highest 
energy state). If the shell is more than half-full (``holes'') this state 
disappears. Otherwise the spectra for particle and hole states are the same 
as the rules of the supersymmetric quantum mechanics 
implies \cite{Balantekin:2007qr}. 

\section{Dynamical Supersymmetries in Nuclear Physics}

Dynamical supersymmetries in nuclear physics start with the algebraic model 
of nuclear collectivity called Interacting Boson Model 
\cite{Armia:1976ky,Arima:1978ha,Arima:1979zz}. 
In this model, 
low-lying quadrupole collective states of even-even nuclei are generated as 
states of a system of bosons occupying two levels, one with angular momentum 
zero ($s$-boson) and one with angular momentum two ($d$-boson). 
It is discussed elsewhere in 
great detail in these proceedings \cite{Casten,Isacker}. 
There are three exactly solvable limits of the simplest form of the 
Interacting Boson Model where neutron and proton bosons are not distinguished:
\begin{itemize}
\item  Vibrational Limit: 
$SU(6) \supset SU(5) \supset SO(5) \supset SO(3)$. 
\item Rotational Limit: 
$SU(6) \supset SU(3) \supset SO(3)$. 
\item Gamma-Unstable Limit:  
$SU(6) \supset SO(6) \supset SO(5) \supset SO(3)$.
\end{itemize}
It is possible to extend the Interacting 
Boson Model to describe odd-even and odd-odd nuclei \cite{Arima:1976zz}. 
Dynamical supersymmetries arise out of certain exactly solvable limits 
of this Interacting Boson-Fermion Model \cite{Iachello:1980av,IachBF}. 

In an odd-even nucleus, in addition to the correlated nucleon pairs 
($s$ and $d$ bosons) we need the degrees of freedom of the unpaired fermions. 
If those unpaired fermions are in the $j_1,j_2,j_3,\cdots$ orbitals then the 
fermionic sector of the theory is represented by the fermionic algebra 
$SU_F(\sum_i(2j_i+1))$ and the resulting $SU(6)_B \times 
SU_F(\sum_i(2j_i+1))$ algebra is embedded in the superalgebra 
$SU(6/\sum_i(2j_i+1))$. In the first example of dynamical supersymmetry 
worked out the unpaired fermion was in a $j=3/2$ (d$_{3/2}$) orbital 
coupled to the nuclei described by the gamma-unstable ($SO(6)$) limit 
of the Interacting Boson Model.  
The resulting $SU(6/4)$ supersymmetry was used  
to describe many properties of nuclei in the Os-Pt region 
\cite{BahaBalantekin:1981kt}. 
Immediately afterwards this supersymmetry was extended to the $SU(6/12)$ 
superalgebra including fermions in s$_{1/2}$, d$_{3/2}$, and d$_{5/2}$ 
orbitals \cite{Balantekin:1982bk}. Theoretical implications of nuclear 
supersymmetries were extensively investigated by many authors 
\cite{Balantekin:1985vq,Schmitt:1989qn,Navratil:1996hp,Barea:2001ra,Leviatan:2003wy,Barea:2005mp}. 
The existence of dynamical supersymmetries in nuclei is 
experimentally established 
\cite{Mauthofer:1986zz,Rotbard:1992hn,Metz:2000sz,Algora:2003yj,Barea:2004ur,Wirth:2004xs}. 

There exists other symmetries of nuclei describing phase transitions between 
different dynamical symmetry limits. For these so-called critical point 
symmetries  
zeros of wavefunctions in confining potentials of the geometric model of 
nuclei give rise to point groups symmetries  \cite{Iachello:2000ye}. 
This scheme is generally applicable to the spectra of systems undergoing 
a second-order phase transition 
between the dynamical symmetry limits $SU(n-1)$ and $SO(n)$. 
The resulting symmetries are either named after the discrete subgroups of the 
Euclidean group $E(5)$, or $X(5)$. A review of the 
experimental searches for critical 
point symmetries at nuclei in the A=130 and A=150 regions is given in 
Ref. \cite{McCutchan:2005ft}.
It is also possible to couple Bohr Hamiltonian of the geometric model 
with a five-dimensional square well potential to a fermion using the 
five-dimensional generalization of the 
spin-orbit interaction. In doing so the E(5) symmetry of 
the even-even nuclei goes into the E(5/4) supersymmetry for odd-even nuclei 
near the critical point \cite{Caprio:2006ez}. Initial experimental tests of 
this latter supersymmetry are encouraging \cite{Fetea:2006vx}

\section{Application of Symmetry Techniques to Subbarrier Fusion} 

In the study of nuclear reactions 
one needs to describe translational motion coupled with internal
degrees of freedom representing the structure of colliding nuclei.  
The Hamiltonian for 
such a multidimensional quantum problem is taken to be
\begin{equation}
\label{95}
H=-\frac{\hbar^2}{2\mu}\nabla^2+V({\bf r})+H_{0}(q)+H_{\rm int} ({\bf r},q),
\end{equation}
where ${\bf r}$ is the relative coordinate of the target-projectile 
pair and ${q}$ represents any internal degrees of freedom of the
system. To study the effect of the structure of the target
nuclei on fusion cross sections near and below the Coulomb barrier one
can take $V(r)$ to be the potential barrier and the third term
in Eq. (\ref{95}) represents the internal structure of the target nuclei.
All the dynamical information about
the system can be obtained by solving the evolution equation
\begin{equation}
\label{96}
i\hbar\frac{\partial\hat U}{\partial t} =
H \hat U
\end{equation}
with the initial condition $\hat U (t_i) =1$. If part of the Hamiltonian 
in Eq. (\ref{95}), say $H_{0}(q)+H_{\rm int} ({\bf r},q)$, is an element 
of a particular Lie algebra, then symmetry methods can significantly 
simplify the solution of the problem. In such a case, if one considers a 
single trajectory ${\bf r}(t)$, the 
$q$-dependent part of the evolution operator becomes an element of the Lie 
group associated with the Lie Algebra mentioned above. (It is possible 
to make this statement rigorous in the context of path-integral formalism 
\cite{Balantekin:1984jv}). 

It is by now well-established that heavy-ion fusion cross-sections below the
Coulomb barrier are several orders of magnitude larger than one would expect
from a one dimensional barrier penetration picture, an enhancement which 
is attributed to the coupling of the translational motion to additional
degrees of freedom such as nuclear and Coulomb excitation, nucleon transfer, or
neck formation \cite{Balantekin:1997yh,Dasgupta:1998yd}. The multidimensional 
barrier penetration problem inherent in subbarrier
fusion can be addressed in the coupled-channels formalism and state of the 
art coupled-channel codes are currently available ({see e.g. 
\cite{Hagino:1999xb}). Although several puzzles remain (such as the large 
values of the surface diffuseness parameter in the nuclear potential 
required to fit the data \cite{Hagino:2003hj,Hagino:2004yd}; 
very steep fall-off of the fusion data for some systems 
at extreme subbarrier energies \cite{Jiang:2004zw}; or
inadequacy of the standard nuclear potentials to simultaneously reproduce
fusion and elastic scattering measurements \cite{Mukherjee:2007iu})
there is overall good agreement between coupled-channels calculations and 
the experimental data. 

An alternative approach is to formulate the problem algebraically by using 
a model amenable to such an approach for describing the nuclear structure,  
such as the Interacting Boson Model \cite{Balantekin:1991em}. 
Not only this approach fits the data well, it can also be used for 
transitional nuclei, the treatment of 
which could be more complicated for coupled-channels calculations 
\cite{Balantekin:1991em,Balantekin:1992qr,Balantekin:1993gy,Balantekin:1997bh}. 

\section{Application of Algebraic Techniques in Nuclear Astrophysics}
    
Nuclear astrophysics has been very successful exploring the origin of 
elements. As our understanding of the heavens evolved, it was realized that 
nuclear data was needed for astrophysics calculations, such as 
nucleosynthesis, stellar evolution, the Big Bang cosmology, x-ray 
bursts, and supernova 
dynamics. Nuclear astrophysics initially started with reaction rate 
measurements, but with the recent 
rapid growth of the observational data, expanding 
computational capabilities, and availability of exotic nuclear 
beams, interest in all aspects of nuclear physics 
relevant to astrophysical phenomena has significantly 
increased \cite{langan}.
In the rest of this section an application of algebraic techniques to a 
nuclear astrophysics problem is presented. 

Light nuclei are formed during the big-bang nucleosynthesis era and 
nuclei up to the iron, nickel and cobalt group are formed during the 
stellar evolution. A good fraction of the nuclei heavier than iron 
were formed in the rapid neutron capture (r-process) nucleosynthesis. 
The astrophysical location of the r-process is expected to be where 
explosive phenomena are present since a large number of interactions are 
required to take place at this location during a
rather short time interval. Core-collapse supernovae which occur following 
the stages of nuclear burning during stellar evolution after the formation of 
an iron core is such a site. 
Neutrino interactions play a very important role in the evolution of
core-collapse supernovae \cite{Balantekin:2003ip}. 
Almost all (99\%) of the gravitational binding energy 
($10^{53}$ ergs) of the progenitor star is 
released in the neutrino cooling of the neutron star formed after the collapse.
It was suggested 
that neutrino-neutrino interactions could play a potentially very significant
role in core-collapse supernovae
\cite{Balantekin:2004ug,Duan:2005cp,Duan:2006jv,Balantekin:2006tg,Hannestad:2006nj}.
A key quantity
for determining the r-process yields is the neutron to seed-nucleus ratio 
(or equivalently neutron-to-proton ratio). Interactions of the neutrinos 
and antineutrinos streaming out of the core both with nucleons and seed nuclei 
determine the neutron-to-proton ratio. Before these neutrinos reach the 
r-process region they undergo matter-enhanced neutrino oscillations as well 
as coherently scatter over other neutrinos. Many-body behavior of this 
neutrino gas is not completely understood, but may have a significant impact 
on r-process nucleosynthesis. 

It was shown that neutrinos moving in matter gain an effective potential due 
to forward scattering from the background particles such as the electrons 
\cite{Wolfenstein:1977ue,Mikheev:gs,Mikheev:wj}.
One can write the Hamiltonian describing neutrino transport in dense matter 
for two neutrino flavors as 
\begin{eqnarray}
H_{\nu} &=& \int dp \left(
\frac{\delta
m^2}{2p}\cos{2\theta} - \sqrt{2} G_F
N_e \right) J_0(p) \nonumber
\\ &+& \frac{1}{2}\int dp \> \frac{\delta m^2}{2p} \> \sin{2\theta}
\left( J_+(p) + J_-(p) \right) , 
\end{eqnarray}
where $N_e$ is the electron density of the medium (assumed to be charge 
neutral and unpolarized), $\theta$ is the mixing angle between neutrino 
flavors, $\delta m^2 = m_2^2 -m_1^2$ is the difference between squares 
of the neutrino masses, 
and $G_F$ is the Fermi coupling strength of the weak 
interactions. 
In the above equation the operators $J^{\pm}, J^0$ has been 
written in terms of 
the neutrino creation and annihilation operators:
\begin{eqnarray}
\label{98}
J_+(p) &=& a_x^\dagger(p) a_e(p), \ \ \ \
J_-(p)=a_e^\dagger(p) a_x(p), \nonumber \\
J_0(p) &=& \frac{1}{2}\left(a_x^\dagger(p)a_x(p)-a_e^\dagger(p)a_e(p)
\right). 
\end{eqnarray}
Here $a_e^\dagger(p)$ and $a_x(p)$ are the creation and annihilation operators 
for the electron neutrino with momentum $p$ and either muon or tau 
neutrino with momentum $p$, respectively. 
The operators in Eq. (\ref{98}) form as many mutually commuting SU(2) 
algebras as the number of allowed values of neutrino momenta:
\begin{equation}
[J_+(p),J_-(q)] = 2 \delta^3(p-q)J_0(p), \ \ \
[J_0(p),J_\pm(p)] = \pm \delta^3(p-q)J_\pm(p) \nonumber
\end{equation}
If, instead of two neutrino flavors, one considers all three active 
neutrino flavors then one should use copies of SU(3) algebras. 
The contribution of neutrino-neutrino forward scattering terms to the neutrino 
Hamiltonian is given by
\begin{equation}
\label{100}
H_{\nu \nu} = \sqrt{2} G_F \int dp \> dq \> (1-\cos\vartheta_{pq})  \>
{\bf J}(p) \cdot {\bf J}(q) ,
\nonumber
\end{equation}
where $\vartheta_{pq}$ is the angle between neutrino momenta ${\bf p}$ and 
${\bf q}$. It is easy to 
include neutrino-antineutrino and antineutrino-antineutrino 
scattering terms in this Hamiltonian \cite{Balantekin:2006tg}, but 
we ignore them here to keep the presentation simple. 

Since a large number of neutrinos $(10^{58})$ are emitted 
during a typical core-collapse it is very difficult to evaluate neutrino 
evolution exactly using the two-body Hamiltonian of Eq. (\ref{100}). 
Instead one uses a mean field approximation where 
the product of two
commuting arbitrary operators $\hat{\cal O}_1$ and $\hat{\cal O}_2$ 
can be approximated as
\begin{equation}
\label{rpa}
\hat{\cal O}_1 \hat{\cal O}_2 \sim \hat{\cal O}_1 \langle \xi |
\hat{\cal O}_2 | \xi \rangle + \langle \xi | \hat{\cal O}_1 |
\xi \rangle \hat{\cal O}_2 -
\langle \xi | \hat{\cal O}_1 |
\xi \rangle
\langle \xi | \hat{\cal O}_2 |
\xi \rangle , \nonumber
\end{equation}
provided that the condition 
\begin{equation}
\label{rpa1}
\langle \xi | \hat{\cal O}_1  \hat{\cal O}_2 | \xi \rangle =
\langle \xi | \hat{\cal O}_1 |
\xi \rangle
\langle \xi | \hat{\cal O}_2 |
\xi \rangle
\end{equation}
is satisfied. In Eqs. (\ref{rpa}) and (\ref{rpa1}) 
One can then replace the exact Hamiltonian of Eq. (\ref{100}) 
with the approximate expression  
\[
H_{\nu \nu} \sim 2\frac{\sqrt{2}G_F}{V} \int dp dq
\> R_{pq} \left( J_0(p) \langle J_0(q) \rangle 
+  \frac{1}{2}
J_+(p) \langle J_-(q) \rangle + \frac{1}{2} J_-(p) \langle J_+(q)
\rangle\right)  
\]
where $R_{pq} =(1-\cos\vartheta_{pq})$ and the averages are calculated over 
the entire ensemble of neutrinos. 
Systematic corrections to this expression are explored in 
\cite{Balantekin:2006tg}.
Calculations using this approximation do not yield conditions 
(i.e. large enough neutron to seed nucleus ratio) 
favorable to 
r-process nucleosynthesis \cite{Balantekin:2004ug}. There is encouraging 
progress in numerical calculations using the exact Hamiltonian of Eq. 
(\ref{100}) \cite{Duan:2005cp,Duan:2006jv}. However an algebraic solution 
to this problem is currently lacking. 

\section{Conclusions}

In many-fermion physics, mean field approaches can describe many 
properties; but are inadequate to describe the whole picture; pairing 
correlations play a crucial role. In fact, pairing correlations are not 
only necessary to understand the structure of rare-earths and actinides, 
but, since they are essential for the description of the neutrino gas 
in a core-collapse 
supernova where many nuclei are produced,  
also necessary to understand 
the existence of such nuclei in the first place. Models exploiting 
symmetry properties and pairing correlations have been very successful. 
These also gave rise to dynamical supersymmetries. 
We showed that using algebraic 
techniques it is possible to solve the s-wave pairing problem almost exactly 
(i.e., reducing it to Bethe ansatz equations) at least 
for a number of simplified cases. 

\acknowledgments

This work was supported in part by the
U.S. National Science Foundation Grant No.\ PHY-0555231 and in part by
the University of Wisconsin Research Committee with funds granted by
the Wisconsin Alumni Research Foundation.

\end{document}